\documentclass[11pt]{article}
\usepackage{epstopdf}
\usepackage[a4paper,hmarginratio=1:1,vmarginratio=2:3,totalwidth=15.3cm,totalheight=22.7cm]{geometry}
\usepackage{bm,epstopdf,epsfig,amsmath,amssymb,amsfonts,colordvi,latexsym,
comment,cancel,verbatim,%extarrows,
slashed}
\usepackage{epsfig,bbm}
\usepackage{graphicx,graphics}
\usepackage[font=md,captionskip=8pt]{subfig}
\usepackage[usenames,dvipsnames]{color}
\usepackage[noadjust]{cite}
\usepackage{xcolor}  %\textcolor{red}{text}
\usepackage{setspace}
 \newcommand{\sg}{\sqrt{g}}    
\newcommand{\sge}{\sqrt{\hat g}}
 \newcommand{\sgh}{\hat g}  
 \newcommand{\RE}{\hat R}  
\newcommand{\LE}{\hat L}  
 \newcommand{\VE}{\hat V} 
\newcommand{\w}{\omega}

\usepackage[utf8]{inputenc}
\allowdisplaybreaks

\newcommand{\cL}{{\cal L}}
\newcommand{\cM}{{\cal M}}

\newcommand{\cO}{{\cal O}}

\newcommand{\ra}{\rightarrow}

\newcommand{\be}{\begin{equation}}
\newcommand{\ee}{\end{equation}}
\newcommand{\bea}{\begin{eqnarray}}
\newcommand{\eea}{\end{eqnarray}}

\newcommand{\baa}{\begin{array}}
\newcommand{\eaa}{\end{array}}

\long\def\symbolfootnote[#1]#2{\begingroup
\def\thefootnote{\fnsymbol{footnote}}\footnote[#1]{#2}\endgroup}
\begin{document} 

\begin{flushright}
\end{flushright}

\bigskip\medskip

\thispagestyle{empty}

\vspace{2.cm}

\begin{center}

  {\Large {\bf  Weyl gauge symmetry and its spontaneous breaking 

\bigskip
 
in Standard Model and inflation}}

\vspace{1.cm}

 {\bf D. M. Ghilencea$\,^{a}$}
and
{\bf Hyun Min Lee$\,^{b,c}$}
\symbolfootnote[1]{E-mails: dumitru.ghilencea@cern.ch, hminlee@cau.ac.kr}

\bigskip
{\small \it $^a$Department of Theoretical Physics, National Institute of Physics
 
and Nuclear Engineering  (IFIN) Bucharest\, 077125, Romania}

{\small \it $^b$Department of Physics, Chung-Ang University, Seoul 06974, Korea}

{\small \it $^c$School of Physics, Korea Institute for Advanced Study, Seoul 02455, Korea}

\end{center}

\medskip
\begin{abstract}
\noindent
We discuss the local (gauged) Weyl  symmetry and its spontaneous breaking and apply it 
to model  building beyond the Standard Model (SM) and inflation. 
In  models with  non-minimal couplings of the scalar fields to the Ricci scalar,
that are conformal invariant,  the spontaneous generation   by a scalar field(s) vev 
 of a  {\it positive}  Newton constant demands a  
 {\it negative}  kinetic term for  the scalar field, or vice-versa. 
This  is naturally avoided in models with additional Weyl gauge symmetry.
The Weyl gauge field $\w_\mu$ couples to the scalar sector but not to the fermionic sector
 of a SM-like  Lagrangian.
The field $\w_\mu$  undergoes  a Stueckelberg  mechanism and  becomes massive 
after ``eating''   the (radial mode) would-be-Goldstone  field (dilaton $\rho$) in the scalar sector.
Before the decoupling of $\w_\mu$,  the dilaton can act as  UV regulator and maintain the Weyl 
symmetry at the {\it quantum} level, with  relevance for  solving the hierarchy problem. 
After the decoupling of $\w_\mu$,  the scalar potential depends only on the remaining (angular 
variables) scalar fields, {that can be the Higgs field, inflaton, etc}.
 We show that a successful  inflation is then possible
with  one of these scalar  fields identified as the inflaton. 
 While our approach is derived in the Riemannian geometry with $\w_\mu$ introduced to avoid ghosts,
the natural framework is  that 
of Weyl geometry which for the same matter spectrum is  shown to generate  the same 
Lagrangian, up to a  total derivative.

\end{abstract}

\newpage

\section{Introduction}

In this letter we discuss the Weyl gauge symmetry and its spontaneous breaking 
together with its implications for model building beyond the Standard Model (SM)
and for inflation. 

One phenomenological motivation  relates to the observation  that the SM with a Higgs mass 
parameter set to zero has a classical scale symmetry  \cite{Bardeen}. 
If this symmetry is preserved at the quantum level 
 by  (a scale-invariant) UV regularisation as in \cite{Englert,S1,Armillis,Monin,D1,Tamarit}, and
 is broken spontaneously only,  it can naturally protect {\it at the quantum level} a 
hierarchy of  fields vev's  of the  theory \cite{S1,R1,D1,K1,D2}.  
The hierarchy  we refer to is that between  the Higgs field vev (electroweak scale) and  that
of ``new physics'' represented  by the vev of  the flat direction  (dilaton) 
associated   with global scale symmetry breaking. Such  hierarchy of vev's 
can be generated  by a classical hierarchy of the dimensionless  
couplings of the theory \cite{Allison,CW-BL}.

A proper study of the hierarchy  problem, based on the above idea,
 demands  including gravity and  generating
spontaneously the Planck scale ($M_p$). This can be  done   in Brans-Dicke-Jordan
theories of gravity \cite{BDJ} 
via a non-minimal  coupling between a scalar field(s) and  the scalar 
curvature ($R$), when this field(s) develops a non-zero vev.   However, demanding the 
theory be conformal invariant and spontaneous-only  breaking of the conformal symmetry, 
leads to  a negative kinetic term for the corresponding scalar field, 
a ``nuisance'' that is  often  quietly glided over. 
{This problem is automatically avoided  in models with 
Weyl gauge symmetry \cite{Weyl,Dirac} and motivated our study of this symmetry 
in Sections~\ref{WI} and \ref{3}.}

{The Weyl gauge symmetry }
is  the natural extension  for conformal invariant models e.g. 
\cite{Weyl,Dirac,Scholz,Scholz2,BDJ,Turok,Smolin,Cheng,Ni,
Moffat1,TW,Davidson,Moffat2,Heisenberg,Oh,Wheeler,Quiros,FRH0,FRH2,FRH1};
the conformal transformation of the metric is extended by the associated
gauge transformation of a Weyl gauge field ($\w_\mu$) which is of geometric origin.
Section~\ref{3} discusses how
 $\w_\mu$  undergoes a  Stueckelberg mechanism and becomes massive by ``eating'' the
would-be-Goldstone field (dilaton $\rho$); here, the dilaton is the radial direction in the field
space of scalar fields ($\phi_j$)  of different non-minimal couplings $\xi_j$ to $R$.
The  Weyl gauge symmetry is then spontaneously broken and 
 there are no negative kinetic terms in the theory. The  vacuum expectation value 
$\langle\rho\rangle$ of the flat direction  (dilaton)  controls the mass of $\w_\mu$ 
and $M_p$. After $\w_\mu$ decouples, the potential depends only on the remaining 
angular variables scalar fields which can account for the Higgs 
field, inflaton, etc\footnote{
The mechanism of the Weyl gauge symmetry breaking used here differs  from that 
in \cite{Oh} where  a complex scalar is considered rather the (real) dilaton,
and  a  Coleman-Weinberg mechanism (unitary gauge)
is used (instead of a Stueckelberg mechanism), 
which  breaks explicitly  the Weyl symmetry by  UV regularization.
An explicit (classical) breaking was also considered in \cite{Davidson} 
for one scalar field case.
}.
Our analysis  extends previous studies \cite{Smolin,Cheng,Ni,Moffat1,Oh}
to  multiple scalar  fields ($\phi_j$) and different non-minimal  couplings ($\xi_j$).

We also  show (Section~\ref{3.4})  how prior to  this symmetry breaking 
the dilaton can enforce a  UV regularization of the quantum corrections that keeps manifest
the Weyl symmetry.  In Weyl-invariant  models the dilaton 
replaces the subtraction scale, thus  maintaining this symmetry at the quantum 
level \cite{Englert,S1,Armillis,Monin,D1,Tamarit}, after which is 
``eaten'' by  $\w_\mu$ and disappears from the spectrum.
One is left with the potential for angular variables fields (e.g. Higgs field, etc).
 This is relevant for the  hierarchy problem in Weyl-symmetric theories.

While our analysis  (Section~\ref{2.2}) is formulated in Riemannian geometry (RG) 
extended by the Weyl gauge  symmetry, the natural framework for this study is   
Weyl conformal geometry (WG) \cite{Weyl,Dirac,Scholz}. In the RG case, imposing the Weyl 
symmetry {to avoid generating ghosts}, leads  to a SM-like  Lagrangian with the 
corresponding current $K_\mu=\partial_\mu K$ where $K_\mu$ interacts with the field  
$\w_\mu$  and $K=\rho^2$.  We show 
that this  Lagrangian  is identical, up to a total derivative 
term, to the simplest Lagrangian one can build in the Weyl geometry for the same set of
 matter fields, using  the curvature  scalar and curvature tensors of WG 
(Section~\ref{2.3}). 
This equivalence  is an  interesting  result that
 follows from the relation between $R$ computed in 
Riemannian geometry  and its 
counterpart $\tilde R$ computed in  Weyl geometry.

We also verify that in the Lagrangian $\cL$ of the SM endowed  with Weyl gauge symmetry,
unlike the Higgs sector, gauge bosons and  fermions do not couple 
to $\w_\mu$ \cite{Moffat1}  
(except a possible kinetic mixing of $\w_\mu$ to  $U(1)_Y$).
$\cL$ can be used for further  phenomenological studies of the Weyl gauge symmetry.

For the case of two scalar fields present
with non-minimal couplings, after the Weyl field $\w_\mu$ decouples, 
the  potential depends only on the angular field $\theta$ and  becomes constant for 
large  $\tan\theta$. We show that  successful inflation is then possible,
in which the  field  $\theta$ is  playing the role of the inflaton.
This is another  result of this work, discussed in Section~\ref{4}.
Our conclusions are presented in Section~\ref{5}.

\section{Implications of  Weyl gauge symmetry} \label{WI}

We review how  models invariant under conformal transformations  become ghost-free 
 while generating spontaneously a positive Newton constant, when a Weyl 
 gauge transformation is added. The Lagrangian so obtained is then shown to be
equivalent to that derived 
 in Weyl geometry,  up to a total derivative; a SM-like model with this symmetry is 
  also constructed.

\subsection{Weyl symmetry or how to obtain  a Lagrangian without ghosts}\label{2.1}

Consider a (local) conformal  transformation of the metric\footnote{
Conventions: metric $(+,-,-,-)$,
$R^\lambda_{\mu\nu\sigma}=
\partial_\nu \Gamma^\lambda_{\mu\sigma}
-\partial_\sigma\Gamma^\lambda_{\mu\nu}
+ \Gamma^\lambda_{\nu\rho}\,\Gamma^\rho_{\mu\sigma}
-\Gamma^\lambda_{\sigma\rho}\,\Gamma^\rho_{\mu\nu}$,
 $R_{\mu\sigma}=R^\lambda_{\mu\lambda\sigma}$, $R=g^{\mu\nu} R_{\mu\nu}$.}  and of a scalar field 
$\phi$ and  a  fermion $\psi$, as follows
\bea
\label{ct}
g_{\mu\nu} &\ra&  g_{\mu\nu}^\prime =  e^{2\,\alpha(x)}\, g_{\mu\nu},
\nonumber\\[3pt]
\phi &\ra& \phi^\prime   =  e^{-\alpha(x)\,\Delta_s}\,\phi, 
\quad\quad \psi \,\,\,\rightarrow\,\,\, \psi' = e^{-\alpha(x)\,\Delta_f}\,\psi.
\eea
% \medskip\noindent 
Then $g^{\mu\nu \prime}= e^{-2\alpha(x)} g^{\mu\nu}$ and
$\sqrt{g^\prime}\!=\! e^{4\alpha(x)}\! \sg$ with
$g\!=\!\vert\det g_{\mu\nu}\vert$.
Here  $\Delta_s\!=\!1$  and $\Delta_f=3/2$.

We would like to generate the Planck scale spontaneously, from the vev of a scalar field $\phi$.
To this purpose one uses that the Lagrangian
%\medskip
\bea\label{d1}
L_1^\pm=\pm \sg\, \frac{\xi}{2}\,
 \Big\{\,\frac{1}{6}\, \phi^2\,R + g^{\mu\nu} \partial_\mu\phi\partial_\nu\phi\,\Big\}
\eea

\medskip\noindent 
 is invariant under transformation (\ref{ct})\footnote{To see this, one uses 
that under eq.(\ref{ct}) $R$ transforms as 
$R\ra R^\prime= e^{-2\alpha(x)} \big( R-6 \, e^{-\alpha(x)} \Box e^{\alpha(x)}\big)$}.
 $\xi$ is the non-minimal coupling and we assume  $\xi\!>\!0$.

Then one is facing the following issue. To   generate  the Einstein term 
\bea
L_E= -\frac 12\, \sg\, M_p^2 \,R
\eea
after spontaneous breaking of conformal symmetry from a vev of $\phi$ from
the first term in (\ref{d1}), one must  take the minus sign in front of (\ref{d1});  that 
means  a negative kinetic term for $\phi$ (ghost) is present in the theory, which 
may not be acceptable.
 Alternatively a positive kinetic term leads to $M_p^2\!<\!0$.  
One usually sets $M_p=\langle\phi\rangle$ (``gauge fixing'' the Planck scale)  
and the ghost  presence is then ignored. Yet, one cannot have the benefit of  conformal symmetry 
 but ignore this ``side effect'', therefore  we would like 
to understand its meaning.

To avoid this problem, we associate to transformation (\ref{ct}) that of a 
(Weyl) vector field $\w_\mu$ \cite{Weyl} 
which, in the light of (\ref{ct}), is of geometric origin
\bea\label{ct2}
\w_\mu\ra \w_\mu^\prime &=&\w_\mu-\frac{2}{q}\,\partial_\mu\alpha(x),
\eea
%
%\medskip\noindent
then consider adding the kinetic term below, with a suitable normalization coefficient
\medskip
\bea
L_2=\frac{1}{2} \,(1+\xi)\,\sg\, g^{\mu\nu} \,\tilde D_\mu \phi\,
\tilde D_\nu \phi, \qquad\qquad \tilde D_\mu\equiv \partial_\mu-\frac{q}{2} \,\w_\mu.
\eea

\medskip\noindent
$L_2$ is  invariant under (\ref{ct}), (\ref{ct2}) since 
$\tilde D_\mu\phi\ra e^{-\alpha}\tilde D_\mu\phi$, due to the presence of $\w_\mu$.
Since $L_1^\pm$ is also invariant under (\ref{ct}), (\ref{ct2}), the sum 
$L_1^\pm\!+\!L_2$  is also invariant. Hereafter  we take  $L_1^-$.
One has
$L_1^-\!+\!L_2=(1/2)\, g^{\mu\nu}\, \partial_\mu\phi\,\partial_\nu\phi 
-(1/12)\,\xi\, \phi^2 R+\cdots$, 
with a canonically normalized kinetic term for $\phi$.
 Thus,  the Planck (mass)$^2$ generated by $\langle\phi^2\rangle$ and  the 
 kinetic term of $\phi$ can be simultaneously positive\footnote{This 
 is automatic in Weyl geometry, see  Section~\ref{2.3} and 
\cite{Smolin,Oh,Moffat1}.}.
 This is made possible by the additional presence of the
 Weyl  field $\w_\mu$; this is a sufficient condition for the  
consistency of the theory (absence of ghosts).

\subsection{SM Lagrangian with Weyl gauge symmetry }\label{2.2}

We use the above observation about $L_1^-+L_2$  to construct a Lagrangian
without ghosts and invariant under eqs.(\ref{ct}), (\ref{ct2}).
For generality,  consider a version of $L_1^-+L_2$ with  more scalar fields $\phi_j$ of non-minimal 
couplings $\xi_j$, then a Weyl-invariant Lagrangian is
\medskip
\bea
L= \sg\,\Big\{ - \frac{\xi_j}{2} \, \Big[\frac{1}{6}\, \, \phi^2_j\,R+ g^{\mu\nu} \,
\partial_\mu\phi_j\, \partial_\nu\phi_j\Big]
\!+\!(1+\xi_j)\, \frac{1}{2} g^{\mu\nu} \tilde D_\mu\phi_j\,\tilde D_\nu\phi_j
-V(\phi_j)\Big\}.
\eea

\medskip\noindent
A summation is understood over repeated index $j=1,2,3\cdots$.
We also added a potential $V(\phi_j)$  for the scalars $\phi_j$; given the conformal symmetry, 
$V$ is a homogeneous function, so
%\medskip
\bea\label{hf}
V(\phi_j)=\phi_k^4\, V(\phi_j/\phi_k), \qquad k={\rm fixed.}
\eea
%\medskip\noindent
$L$ can be re-written as
\medskip
\bea\label{LL}
L=
\sg\,\Big\{\!-\frac{\xi_j}{12} \,\phi_j^2\,R +\!\frac{g^{\mu\nu}}{2}
(\partial_\mu\phi_j)\,(\partial_\nu\phi_j)
- \frac{q}{4} \,g^{\mu\nu}\, \w_\mu\,K_\nu 
+ \!\frac{q^2}{8}\,K\, \w_\mu\,\w^\mu\!
-V(\phi_j)\! \Big\},
\eea
%\medskip\noindent
where 
%\medskip
\bea
K_{\nu}=\partial_\nu K, \qquad K=(1+\xi_j)\,\phi_j^2.
\eea

\medskip\noindent
 $L$ above is  invariant under (\ref{ct}), (\ref{ct2}), 
for {\it all values} of  $\xi_j$, thanks to the $\w_\mu$-dependent 
terms. $L$ has positive kinetic term for $\phi_j$ and $M_p^2>0$ when generated by the vev of 
$\langle\phi\rangle$ (assuming $\xi_j>0$). 
In the absence of the $\w_\mu$-dependent part,
 $L$ is not  conformal (unless $\xi_j=-1$),
 but only global conformal.
Unlike in gauge theories, $\w_\mu$ is a vector under a {\it real} 
transformation of the  fields $\phi_j$ (missing the $i$ factor). The associated current $K_\mu$ 
 is non-zero for $\phi_j$ reals.

Further, we include a  kinetic term for  $\w_\mu$ with  the ``usual'' (pseudo)Riemannian 
definition 
\bea 
\label{gauge}
L_g=-\,\frac{\sg}{4} \, g^{\mu\rho}\, g^{\nu\sigma}\,F_{\mu\nu}\,F_{\rho\sigma},
\qquad
F_{\mu\nu}= D_\mu \w_\nu -D_\nu \w_\mu,
\quad
 D_\mu \w_\nu=\partial_\mu \w_\nu -\Gamma_{\mu\nu}^\rho \w_\rho,
\eea
$L_g$ is  invariant under (\ref{ct}), (\ref{ct2}), since 
 the metric part is invariant and $F_{\mu\nu}$ ($=\!\partial_\mu\w_\nu\!-\!\partial_\nu\w_\mu$)
is invariant, too.
The Riemann connection\footnote{The Riemann affine connection used here is
$\Gamma_{\mu\nu}^\rho=(1/2)\,g^{\rho\beta} \big[ \partial_\nu g_{\beta\mu} +\partial_\mu g_{\beta\nu}
-\partial_\beta g_{\mu\nu}\big]$.} $\Gamma_{\mu\nu}^\rho$, symmetric in $\mu,\nu$,
 is  not invariant under (\ref{ct}).

Finally, one can consider the Weyl-invariant Lagrangian $L_f$ for 
the massless fermions  of the theory that transform under (\ref{ct}).
$L_f$ has  the usual form in (pseudo)Riemann space
\be\label{fermions}
L_f= \sqrt{g}\,  {\bar\psi}\, i\gamma^a\, e^\mu\,_a  { D}_\mu  \psi,
\qquad
{D}_\mu\psi =\Big(\partial_\mu+\frac{1}{2} \omega^{ab}_\mu \sigma_{ab}
\Big)\,\psi
\ee
where $\omega^{ab}_\mu=e^{\lambda b}(-\partial_\mu e_\lambda\,^a+e_\nu\,^a \Gamma^\nu_{\mu\lambda})$ 
is the spin connection and $\sigma_{ab}=\frac{1}{4}[\gamma_a,\gamma_b]$. 
Note that $g_{\mu\nu}=e_\mu\,^a e_\nu\,^b \eta_{ab}$ and $e^\mu\,_a e_\nu\,^a=\delta^\mu_\nu$. 
Under  a Weyl transformation of the metric, eq.(\ref{ct}), 
the vielbein $e_\mu^a$ transforms as  $e_\mu\,^{a\, \prime}=e^{\alpha(x)} e_\mu\,^a$, 
while for the spin connection   we have
$\omega^{ab\,\prime}_\mu=\omega^{ab}_\mu + (e_\mu\,^a e^{\nu b}-e^{\nu a} e_\mu\,^b)\partial_\mu\alpha$.
Then it can be shown that $L_f$ is  invariant  under a Weyl gauge transformation, 
eqs.(\ref{ct}), (\ref{ct2}), 
and  there is no  coupling of fermions to  the gauge field $\w_\mu$!

Regarding the SM gauge fields kinetic terms ($L_G$), these are invariant under Weyl
gauge symmetry. Indeed, the gauge fields presence under the covariant derivative
that contains $\partial_\mu$ shows that these are invariant, since coordinates
do not transform under (\ref{ct}). 
Therefore, there is no coupling between
SM gauge fields and $\w_\mu$\footnote{An exception 
 is a possible kinetic mixing of the field strength of
$\w_\mu$ to that of $U(1)_Y$ \cite{Hm}.}.
 For example, for the  $U(1)_Y$ gauge field $A_\mu$,
the covariant derivative can be written as 
 $ D_\mu A_\nu=\partial_\mu A_\nu - \Gamma_{\mu\nu}^\rho A_\rho$.
The gauge kinetic terms do not contain the Christoffel symbols
because $F_{\mu\nu}= D_\mu A_\nu-D_\nu A_\mu=\partial_\mu A_\nu-\partial_\nu A_\mu$.

The sum, $\cL=L+L_g+L_f+L_G$, is the total SM-like Lagrangian with
Weyl gauge symmetry\footnote{
One could also add a  Weyl tensor-squared term to the action  which is 
invariant under  (\ref{ct}) or a quadratic term in the Weyl scalar curvature $\tilde R^2$, 
see \cite{new1,new2} for further  details.}
 which is invariant under (\ref{ct}), (\ref{ct2}).  
Here $L$ is immediately adapted to accommodate the Higgs doublet of the SM
with one of the $\phi_j$ fields to account for the Higgs neutral scalar. In conclusion, 
we have a SM-like Lagrangian that is invariant under (\ref{ct}), (\ref{ct2}).

\subsection{From  Riemann to Weyl conformal geometry}\label{2.3}

The presence of the Weyl gauge field in our model 
in  the Riemannian geometry and invariant under (\ref{ct}), (\ref{ct2}) 
is natural in  Weyl's conformal geometry  \cite{Weyl,Dirac} (also \cite{Scholz}).
Following \cite{Dirac} we write the SM-like Lagrangian with this symmetry 
directly in Weyl geometry and we verify that it agrees  with that
 of the previous section, built in the Riemannian geometry 
(with $\w_\mu$ introduced to avoid ghosts).

Weyl geometry is a scalar-vector-tensor theory of gravity and thus 
provides a generalization (to classes of equivalence) of Brans-Dicke-Jordan scalar-tensor theory  
\cite{BDJ}
and of other  conformal invariant models  \cite{Turok}. It was  used for model building 
\cite{Smolin,Cheng}  
with renewed recent interest in  \cite{Moffat1,TW,Davidson,Moffat2,Heisenberg,
Oh,Wheeler,Quiros} and applications to inflation, see e.g.
 \cite{FRH0,FRH2,FRH1,GG1,GG2,GG3,GG4,GG5,higgsdilaton}.
 If the Weyl 
field is set to zero, one obtains  (Weyl integrable) models similar to Brans-Dicke-Jordan
 theory \cite{Quiros}.

In Weyl geometry the curvature scalars and tensors and the connection 
are  different from the Riemannian case where they are induced by the metric alone. 
In Weyl geometry
\medskip
\bea\label{tGamma}
\tilde\Gamma_{\mu\nu}^\rho=
\Gamma_{\mu\nu}^\rho+
\frac{q}{2}\,\Big[ \delta_\mu^\rho \,\w_\nu +\delta_\nu^\rho\,\w_\mu- g_{\mu\nu}\,\w^\rho\Big],
\eea

\medskip\noindent
where $\Gamma_{\mu\nu}^\rho$ are the connection coefficients in the Riemannian geometry.
Under (\ref{ct}), (\ref{ct2}) the coefficients $\tilde\Gamma_{\mu\nu}^\rho$ are invariant, 
as one can easily check. The system is torsion-free.
The Riemann tensor in Weyl geometry is then generated by the ``new'' $\tilde \Gamma_{\mu\nu}^\rho$  
\medskip
\bea
\tilde R^\lambda_{\mu\nu\sigma}=
\partial_\nu \tilde\Gamma^\lambda_{\mu\sigma}
-\partial_\sigma\tilde\Gamma^\lambda_{\mu\nu}
+ \tilde\Gamma^\lambda_{\nu\rho}\,\tilde\Gamma^\rho_{\mu\sigma}
-\tilde\Gamma^\lambda_{\sigma\rho}\,\tilde\Gamma^\rho_{\mu\nu},
\eea

\medskip\noindent
and then $\tilde R_{\mu\sigma}=\tilde R^\lambda_{\mu\lambda\sigma}$, $\tilde R=g^{\mu\nu} \tilde R_{\mu\nu}$.
We can then compute $\tilde R$ and  find
\medskip
\bea\label{tildeR}
\tilde R& =& 
R- 3\,q\,
\Big[\partial_\mu \w^\mu +\frac12 \w^\rho\, g^{\lambda\beta}\,\partial_\rho \,g_{\lambda\beta}\Big]
-\frac32 q^2 \w^\mu \,\w_\mu
\nonumber\\[-4pt]
&=& 
R-3 \,q\,   D_\mu \w^\mu -\frac32 q^2 \, \w^\mu \w_\mu.
\eea
Then  under transformations (\ref{ct}) and (\ref{ct2}),
%\medskip
\bea
\tilde R\ra \tilde R^\prime=e^{- 2\alpha(x)} \tilde R.
\eea
As a result
\bea
 L_{1w}=-\sg\, \frac{1}{12}\,\xi_j\,\phi^2_j \,\tilde R,\qquad \textrm{(sum over $j=1,2.$)},
\eea
%\medskip\noindent
is invariant under combined transformations (\ref{ct}), (\ref{ct2}). 
This is unlike  in the Riemannian case of the previous section
 where  the non-minimal coupling term in the action was not invariant.

Further, we can define a kinetic term for $\phi$ in Weyl geometry,  invariant under (\ref{ct}), (\ref{ct2})
\bea
L_{2w}=\frac{1}{2}\, \sg\, g^{\mu\nu}\, \tilde D_\mu \phi_j \,\tilde D_\nu \phi_j-\sg \,V(\phi_j).
\eea

\medskip\noindent
We also have a gauge kinetic term ($L_{3w}$) for $\w_\mu$, now defined by 
new coefficients $\tilde \Gamma$ of (\ref{tGamma})
\be
L_{3w}=-\frac{\sg}{4} g^{\mu\rho} \,g^{\nu\sigma}\, F_{\mu\nu} F_{\rho\sigma},\qquad
F_{\mu\nu}=\tilde D_\mu \w_\nu-\tilde D_\nu \w_\mu, \qquad
\tilde D_\mu \w_\nu=\partial_\mu\w_\nu-\tilde\Gamma_{\mu\nu}^\rho \w_\rho.
\ee
However,  $\tilde \Gamma_{\mu\nu}^\rho$ are symmetric in $\mu\leftrightarrow\nu$
and also invariant under Weyl transformation eqs.(\ref{ct}), (\ref{ct2}). 
Thus, $F_{\mu\nu}$ and $L_{3w}$ are  equal to their counterparts in 
the previous section, eq.(\ref{gauge}), so $L_{3w}=L_g$.
 The same can be said about the SM gauge  fields kinetic terms. 

Further, the fermionic Lagrangian is defined with the Weyl connection, as follows,
\be\label{fermions2}
L_{4w}= \sqrt{g}\,  {\bar\psi}\, i\gamma^a\, e^\mu\,_a  {\tilde D}_\mu  \psi,
\qquad
{\tilde D}_\mu\psi =\Big(\partial_\mu+\frac{1}{2} {\tilde\omega}^{ab}_\mu \sigma_{ab}
-\frac{3}{4}q\,\omega_\mu\Big)\,\psi
\ee
where ${\tilde\omega}^{ab}_\mu\!=e^{\lambda b}(
-\partial_\mu e_\lambda\,^a\!+\!e_\nu\,^a\, {\tilde\Gamma}^\nu_{\mu\lambda})$. 
However, one shows  \cite{Moffat1} that $L_{4w}\!=L_f$ with $L_f$ of (\ref{fermions}).

Adding together $L_{1w}$, $L_{2w}$, $L_{3w}$, and $L_{4w}$,   
each of these  invariant under (\ref{ct}), (\ref{ct2}), 
we obtain a total  Lagrangian for the case of Weyl geometry.
It is interesting to see that  this Lagrangian is equal to 
$L+L_g+L_f$ of (\ref{LL}), (\ref{gauge}) and (\ref{fermions}), 
up to a total derivative term. This follows from the relation
\medskip\noindent
\bea\label{equiv}
L_{1w}+L_{2w}=L + \frac{q}{4}\, \xi_j\,\partial_\mu\Big[ \sg\, \phi_j^2\,\w^\mu\Big].
\eea

\medskip\noindent
To show eq.(\ref{equiv}), one uses  the relation between $\tilde R$ and $R$ of eq.(\ref{tildeR}) 
that relates Weyl and Riemann scalar curvatures and that
 $\partial_\lambda g=g\,g^{\mu\sigma}\,\partial_\lambda g_{\sigma\mu}$.

Eq.(\ref{equiv})  shows that our model  agrees (for two fields case) with that  in
 \cite{Moffat1}  built within Weyl geometry from the onset and following \cite{Dirac}.
We  thus obtained the same Lagrangian in  Riemann and Weyl geometry, 
albeit with different initial motivations.
Our motivation  for a consistent, ghost-free conformal action, with this symmetry broken spontaneously,
lead us to introduce a gauge transformation and Weyl gauge field associated to (\ref{ct}).

\section{Spontaneous breaking of Weyl gauge symmetry}\label{3}

 In this section we show how the Weyl conformal symmetry of our model is spontaneously broken
for one or more scalar fields of non-minimal couplings $\xi_j$  to $R$.
  Then, we show that the (radial mode)  would-be Goldstone boson (dilaton $\rho$) of the Weyl symmetry 
  decouples from the angular variables  fields   due to a Stueckelberg mechanism for the Weyl 
 gauge field which becomes massive.
Before decoupling, the dilaton can provide a scale-invariant ultraviolet (UV)
regularisation for models in which quantum scale invariance  is important.

\subsection{One scalar field and Stueckelberg mechanism for $\w_\mu$}

Let us first show  how  spontaneous breaking of Weyl symmetry happens for  one
scalar field $\phi$.
Then $L$ of eq.(\ref{LL}) simplifies (no sum over $j$) and we  replace $\phi_j\ra \phi$, then 
\bea
K=(1+\xi)\,\phi^2,\qquad  
V=\frac{\lambda}{4!}\phi^4, \label{onefield}
\eea
where $V$ is the only  one allowed by the Weyl symmetry.
To decouple the scalar field fluctuations  from $R$,  we go to the Einstein frame 
by rescaling the metric to
\bea\label{metric1}
\hat g_{\mu\nu}=\Omega\, g_{\mu\nu}, \qquad
\Omega=\frac{\xi}{6}\,\frac{\phi^2}{\langle\phi\rangle^2}.
\eea

\medskip\noindent
Hereafter a hat on a variable denotes the Einstein frame value of that variable.
From eq.(\ref{LL}) for one field and  eq.(\ref{metric1}) we  obtain the tensor-scalar part of 
Einstein-frame Lagrangian as
\medskip
\bea
\LE= \sge\, \Big\{-\frac{1}{2}\,\langle\phi\rangle^2\,\RE
+\frac{3}{4}\langle\phi\rangle^2 (\partial_\mu \ln\Omega)^2
+\frac{1}{\Omega}\,\Big[\,\frac{1}{2}(\partial_\mu\phi)^2
+ \frac{q^2}{8}K \w_\mu\w^\mu 
-\frac{q}{4} \w^\mu K_\mu \Big] 
-\frac{V}{\Omega^2}\Big\} 
\eea

\medskip\noindent
giving
\bea
\LE &= & \sge\, \Big[ -\frac{1}{2}\langle\phi\rangle^2 \RE
+3 \langle\phi\rangle^2 \Big(1+\frac{1}{\xi}\Big)\Big(\frac{\partial_\mu\phi}{\phi}\Big)^2  
+\frac{3}{4}\,q^2\langle\phi\rangle^2\Big(1+\frac{1}{\xi}\Big)\w_\mu\,\w^\mu 
 \nonumber \\
&&\quad 
-3\, q\,\langle\phi\rangle^2\Big(1+\frac{1}{\xi}\Big) \w^\mu\,{\partial_\mu\ln \phi}
-\frac{3\lambda}{2\,\xi^2} \langle\phi\rangle^4  \Big].
\eea

\medskip\noindent
where all contractions are with the new metric $\hat g_{\mu\nu}$. Finally, we introduce
\medskip
\be
\w^\prime_\mu =\w_\mu -\frac{2}{q} \partial_\mu\ln\phi. 
\label{weyl1}
\ee 
giving
\bea
\LE= \sge\, \Big[-\frac{1}{2}\langle\phi\rangle^2\,\RE
+\frac{3}{4}\, q^2\,\langle\phi\rangle^2\,
\Big(1+\frac{1}{\xi}\Big)\,\w^\prime _\mu \w^{\prime\mu}
-\frac{3\lambda}{2\,\xi^2}\langle\phi\rangle^4  \Big]
\eea

\medskip\noindent
As a result, the scalar (dilaton) field $\phi$   
is ``eaten'' by the Weyl gauge boson $\w_\mu$. The  mass of $\w_\mu$ is
$m_\w^2=(3 q^2/2)(1+1/\xi)\langle\phi\rangle^2$. Therefore, conformal symmetry is broken 
spontaneously as in the Stueckelberg formulation for a massive $U(1)$ {\it without} a 
corresponding Higgs mode.
The number of degrees of freedom remains the same (three): in Jordan frame we had
a real scalar  and a massless vector, while in Einstein frame, after breaking 
there is no scalar field but a massive vector boson.
Also note that the gauge kinetic term $L_g$ of $\w_\mu$, see eq.(\ref{gauge}),
 is invariant under (\ref{metric1}), (\ref{weyl1}).
The scalar potential becomes a cosmological 
constant, $V_0=3\lambda\langle\phi\rangle^4/(2\xi^2)$, in Einstein frame.

Transformation (\ref{weyl1}) may be seen as 
a Weyl gauge transformation (\ref{ct2}) with $\alpha=\ln \sqrt{\Omega}$
corresponding to  (\ref{metric1}).
Then the scalar field $\phi$ transforms  according to eq.(\ref{ct}) into
\medskip
\bea
 \phi^\prime = e^{-\ln\sqrt\Omega}\phi
=\sqrt{6/\xi}\,\, \langle\phi\rangle,
\eea

\medskip\noindent
so $\phi^\prime$ is not dynamical anymore. 
Therefore spontaneous breaking of conformal symmetry fixing 
 the Planck scale (to $M_p=\langle\phi\rangle$) and  Stueckelberg mechanism
are related to  a Weyl transformation to a special ``unitary'' gauge 
(``gauge fixing'').

 While we  used
$\langle\phi\rangle\!\not=\!0$ in the definition of $\Omega$ and subsequent equations,
 this is actually not needed and an arbitrary  mass scale $\cM$
can be used instead of $\langle\phi\rangle$, corresponding to a
 different ``gauge fixing'' (and different Planck scale!). 
Indeed, Stueckelberg mechanism is 
a re-arrangement of the degrees of freedom (that does  not
 require $\langle\phi\rangle\not=0$).
Using an arbitrary $\cM$ 
 is consistent with the fact that for a single scalar field
in a Weyl-invariant theory $\langle\phi\rangle$ cannot 
be determined from  the condition  $4\,V(\phi)-\phi V^\prime(\phi)=0$
 which is automatically respected, hence $\langle\phi\rangle$ remains a parameter (unknown). 
This condition is also related to the conservation of the current $D^\mu K_\mu=0$, which
 for  a FRW metric leads to a constant  solution $\langle\phi\rangle$ \cite{FRH0,FRH2} that is
 not fixed by the theory.

\subsection{Two scalar fields and Stueckelberg mechanism for $\w_\mu$}
\label{section3.2}

Let us consider now the more interesting case of two scalar fields in eq.(\ref{LL}), (j=1,2).
Then
\medskip
\be\label{two}
K\!=(1+\xi_1)\,\phi_1^2+(1+\xi_2)\,\phi_2^2,
\ee

\medskip\noindent
Since  $V$ is a homogeneous function of fields, one can have
\medskip
\be\label{Vh1}
V(\phi_1,\phi_2)=\frac{\lambda_1}{4!}\phi_1^4 
+\frac{\lambda_{12}}{4}\phi_1^2\,\phi_2^2
+\frac{\lambda_2}{4!}\,\phi_2^4.
\ee
In particular, if \, $3 \lambda_{12}=-\sqrt{\lambda_1\lambda_2}$, then
\be\label{Vh2}
V(\phi_1,\phi_2)=\frac{\lambda_1}{4!}\,\Big(
\phi_1^2 -\frac{\sqrt \lambda_2}{\sqrt\lambda_1}\,\phi_2^2\Big)^2.
\ee

\medskip\noindent
$V$ can also contain terms like $\phi_1^6/\phi_2^2$, etc \cite{D1}.
The results below are for a  general homogeneous function $V(\phi_{1,2})$
i.e. it has a flat direction:
 $V(\phi_{1,2})=(\phi_1^2-k_0\, \phi_2^2)^2\,f(\phi_1/\phi_2)$, ($k_0$=constant).

To decouple  $R$ from the fluctuations of $\phi_{1,2}$, we consider a transformation
to the Einstein frame.
Let us   perform a metric rescaling of $L$ eq.(\ref{LL}),  to 
\medskip
\bea\label{Omega2fields}
\sgh_{\mu\nu}=\Omega\, g_{\mu\nu},\qquad
\Omega=\frac{1}{6\,v^2}\,(\xi_1\phi^2_1+\xi_2\phi^2_2), 
\qquad v^2\equiv\langle\xi_1\phi_1^2+\xi_2\phi_2^2\rangle.
\eea

\medskip\noindent
Here $v$ ensures that $\Omega$ is dimensionless\footnote{{
As for the one-field case we could use instead of $v$ an arbitrary mass scale.}}. 
From eq.~(\ref{LL}) for two fields and  with (\ref{Omega2fields}),
 we obtain the corresponding  Einstein-frame Lagrangian as
\medskip
\bea
\LE\!&=&\!\! \sge \,\Big[ -\frac{1}{2}\,v^2\,\RE 
+\frac{3}{4}\,v^2\,(\partial_\mu \ln\Omega)^2
\nonumber\\
&+&\frac{1}{\Omega}\,
\Big(\,\frac{1}{2}(\partial_\mu\phi_1)^2
+\frac{1}{2}(\partial_\mu\phi_2)^2 
+ \frac{q^2}{8}K\, \w_\mu\w^\mu 
-\frac{q}{4} \w^\mu K_\mu \Big)
-\VE
\Big] 
\eea

\medskip\noindent
where all contractions are with the new metric $\sgh_{\mu\nu}$;  $\Omega$, 
 $K$ and $\VE$ are  functions of $\phi_{1,2}$ with
\medskip
\bea\label{VE}
\VE(\phi_1,\phi_2)=\frac{1}{\Omega^2} \,V(\phi_1,\phi_2).
\eea
Then
\bea
\LE
&=& \sge\, \Big[ -\frac{1}{2}\,v^2\,\RE 
+\frac{1}{2}G_{ij}\, \partial_\mu\phi_i \,\partial^\mu \phi_j 
+\frac{q^2}{8} \frac{K}{\Omega} \w_\mu\w^\mu
-\frac{q}{4}  \w^\mu \frac{K_\mu}{\Omega}-\VE  \Big]
\label{twofield2}
\eea
%\medskip\noindent
where
%\medskip
\bea\label{Gmatrix}
G_{ij} =
\frac{1}{6\,v^2\,\Omega^2}
\left(\begin{array}{cc} 
\xi_1(1+\xi_1)\phi^2_1
+\xi_2\phi^2_2 &  \xi_1 \xi_2\, \phi_1\phi_2 \\   
 \xi_1 \xi_2\, \phi_1\phi_2 &  \xi_2(1+\xi_2)\phi^2_2+\xi_1\phi^2_1 
\end{array}\right), \quad  i,j=1,2.
\eea

\medskip\noindent
The kinetic terms in $\LE$ become diagonal (no mixing)
 in a new  fields basis  of ($\rho$, $\theta$) where
\medskip
\bea\label{phi1}
\phi_1&=& \frac{1}{\sqrt{1+\xi_1}}\,\rho\,\sin\theta, \nonumber\\[-2pt]
\phi_2&=& \frac{1}{\sqrt{1+\xi_2}}\, \rho \,\cos\theta.
\eea

\medskip\noindent
It is more  illustrative however to first bring  the Weyl terms in $\LE$ to a
quadratic form using
\medskip
\bea\label{om}
\w'_\mu=\w_\mu -\frac{1}{q}\partial_\mu\ln  K,
\eea

\medskip\noindent
where notice that   $K=\rho^2$.
Adding $\hat L_g$ of eq.(\ref{gauge}) with eq.(\ref{Omega2fields})  for the Weyl field $\w_\mu$
 then
\medskip
\bea
\LE+\hat L_g\!\!\!& = &\!\!\!  \sge \,\Big[-\frac{1}{2}\,v^2\,\RE 
+\frac{1}{2}\,G_{ij} \partial_\mu\phi_i \,\partial^\mu \phi_j 
-\frac{1}{8K \Omega} (\partial_\mu K)^2
-\frac{1}{4} F'_{\mu\nu}F^{\prime \mu\nu}
+ \frac{K}{8\Omega}\, q^2 \w'_\mu \w^{\prime \mu} 
- \VE\Big] 
\nonumber \\
&=&\!\! \sge \, \Big[-\frac{1}{2}\,v^2\,\RE 
+\frac{1}{2}\, T_{ij}\, \partial_\mu\phi_i \partial^\mu \phi_j
-\frac{1}{4} F'_{\mu\nu}F^{\prime \mu\nu}
+ \frac{K}{8\Omega}\, q^2 \w'_\mu \w^{\prime \mu} 
-\VE \Big]  \label{twofield3}
\eea

\medskip\noindent
where  $F'_{\mu\nu}=\tilde D_\mu \w'_\nu-\tilde D_\nu\w'_\mu$ 
is invariant under (\ref{om}). Above we denoted   $T_{ij}=G_{ij}+H_{ij}$, ($i,j=1,2$),  with:
\medskip
\be\label{Hmatrix}
H_{ij} =
% \frac{-6\,v^2}{\xi_1\phi^2_1\! +\xi_2\xi^2_2} 
-\frac{1}{\Omega}
% \,\frac{1}{(1+\xi_1)\phi^2_1\! +(1+\xi_2)\phi^2_2}
\, \frac{1}{K}
%\nonumber \\[4pt]
% &&\qquad\qquad\times\,\,  
\left(\!\begin{array}{cc}
 (1+\xi_1)^2\phi^2_1 & (1+\xi_1)(1+\xi_2)\phi_1 \phi_2
 \\ 
(1+\xi_1)(1+\xi_2)\phi_1 \phi_2  & (1+\xi_2)^2 \phi^2_2  
\end{array}\! \right).
\ee

\medskip\medskip\noindent
In the new basis (\ref{phi1}) the scalar kinetic terms in 
eq.~(\ref{twofield3}) are reduced to a single term and
\medskip
\bea\label{Lfinal}
\LE+\hat L_g=\sge \,\Big[ \frac{-1}{2} \,v^2\,\RE 
+\frac12\,
F(\theta)\,
%\frac{6\,b}{\xi_2} \frac{\tan^2\theta+a}{(\tan^2\theta+b)^2} 
\, v^2\,(\partial_\mu\tan\theta)^2
-\frac{1}{4} \,F'_{\mu\nu} F^{' \mu\nu}+ \frac{1}{2} m^2(\theta) \,\sgh^{\mu\nu}  w_\mu' w_\nu'
-\VE\Big]
\eea
with
\bea
F(\theta)=\frac{6\,b}{\xi_2} \frac{\tan^2\theta+\xi_2/\xi_1}{(1+\tan^2\theta)(\tan^2\theta+b)^2},
\qquad
% a=\frac{\xi_2}{\xi_1}, \quad\quad 
 b= \frac{\xi_2(1+\xi_1)}{\xi_1(1+\xi_2)}. 
%
% \xi_1'=\frac{\xi_1}{1+\xi_1},  \,\,\, \xi_2'=\frac{\xi_2}{1+\xi_2}.
\eea

\medskip
Therefore,  we are  left with the ``angular'' kinetic term for $\theta$ only.
The kinetic term of the  radial (Goldstone) coordinate $\rho$\, (where $\rho^2= K$)
has disappeared, via Stueckelberg mechanism, as it was ``eaten'' by  the Weyl gauge 
 boson $\w'$ in  eq.(\ref{om}). This is  similar to the case with one scalar field in 
eq.~(\ref{weyl1}). Thus, in the Einstein frame we  have a massive vector boson and one 
(real) scalar field left ($\theta$), while in Jordan frame we had two (real) scalar fields 
and a massless $\w_\mu$, so the number of degrees of freedom is again conserved.

Further, the function $m^2(\theta)$ in (\ref{Lfinal}) is given by
\medskip
\bea\label{thetaf}
m^2(\theta)=\frac{q^2\, K}{4\Omega}=
 \frac{3\,q^2}{2}\, \frac{v^2 \,(1+\xi_1) (1+\xi_2)(1+\tan^2\theta)}{
\xi_1 (1+\xi_2)\,\tan\theta^2  +\xi_2\,(1+\xi_1)},
\eea
with 
\bea\label{vvv}
v^2=
\langle\rho\rangle^2 \Big[
\frac{\xi_1}{1+\xi_1}
\sin^2\langle\theta\rangle
+ \frac{\xi_2}{1+\xi_2} 
\cos^2\langle\theta\rangle\Big].
\eea

\medskip\noindent
Notice that if $\xi_1=\xi_2$ or if  $\tan\theta$ is large, 
 the function $m^2(\theta)$ is actually independent
of $\theta$ and then the  Weyl gauge field ($\w_\mu^\prime$) and the field $\theta$  decouple.

On the ground state $\theta=\langle\theta\rangle$ and the mass of $\w_\mu$ is
\bea\label{mass}
m^2(\langle\theta\rangle) =\frac{3}{2}\,q^2 \langle\rho\rangle^2.
\eea
% \medskip\noindent
The  mass of  $\w_\mu$ is thus determined by $\langle\rho\rangle$ alone; unlike $\theta$
whose vev is determined from $\VE$ (see below), 
$\langle\rho\rangle$  cannot be predicted  by the theory and is a free 
parameter (flat direction)\footnote{$\langle\rho\rangle$
may be fixed by quantum corrections; however, in quantum scale invariant theories
only ratios of field vev's (scales) can be determined (in terms of dimensionless
couplings), so it remains a free parameter.}.

The Planck  scale $M_p^2=v^2$, eq.(\ref{vvv}),  depends in general on $\langle\theta\rangle$.
This is not a problem since unlike $\rho$, the field variable $\theta$ does not change under a
Weyl transformation, eq.(\ref{ct}). However, if the theory has an $\cO(2)$ symmetry, i.e. 
identical non-minimal couplings  $\xi_1=\xi_2$,
 then $M_p$ is determined by the vev of the dilaton alone 
$M_p^2=v^2=\xi_1\langle\rho\rangle^2/(1+\xi_1)$; in this case, the would-be Goldstone 
(dilaton) field  $\rho$ ``eaten'' by $\w_\mu$  and ``fixing'' its mass also  fixes
the Planck scale. The same is true in the limit of 
large $\tan\theta\ra \infty$,  when $M_p^2=v^2=\xi_1\langle\rho\rangle^2/(1+\xi_2)$.

Regarding the potential $\VE$  in eq.(\ref{Lfinal}), it is given by eq.(\ref{VE})
expressed in terms of the new field variables $\rho,\theta$. With eq.(\ref{hf}) 
and  $V(\phi_1, \phi_2)$ the initial potential in Jordan frame, then
\medskip
\bea
\VE= 36\, v^4\,\frac{b^2}{\xi_2^2}\, \frac{V(c\,\tan\theta,1)}{(\tan^2\theta+b)^2},
\qquad \textrm{where}\qquad
c=\sqrt{\frac{1+\xi_2}{1+\xi_1}},
\eea

\medskip\noindent
which depends on $\theta$ only. 
Finally, another mechanism to decouple the dilaton was studied in \cite{FRH0}  
using  a  global version of  the  Weyl  symmetry studied here
(and assuming  this survives black hole physics \cite{Banks}).

\subsection{More fields and Stueckelberg mechanism}

The  Stueckelberg mechanism for $\w_\mu$ can be extended for  more
scalar fields with non-minimal couplings,  using general coordinates. For
three fields 
$\phi_1=(1/\sqrt{1+\xi_1})\rho\sin\theta\cos\zeta$, 
$\phi_2=(1/\sqrt{1+\xi_2})\rho\sin\theta\sin\zeta$, 
$\phi_3=(1/\sqrt{1+\xi_3})\rho\cos\theta$. 
As before, the kinetic term of radial field $\rho$ is the Goldstone  eaten by the 
vector boson $\w_\mu$ of mass $q^2 K/(4\Omega)\vert_{\theta=\langle\theta\rangle}$.
 One is left with  kinetic terms
for the angular-coordinates fields $\theta$, $\zeta$; similarly, the scalar potential 
will depend only on these fields. This generalization is useful in cases where
one of the scalar fields left is a Higgs field, while the other is a second
Higgs-like scalar, inflaton, etc. The scalar potential is then
\medskip
\bea
\VE(\theta,\phi)=\frac{1}{\Omega^2} \,V(\phi_1,\phi_2,\phi_3)=
\frac{36\,v^4\,V(z_1,z_2,z_3)}{(\xi_1 z_1^2+\xi_2 z_2^2+\xi_3 z_3^2)^2}
\eea

\medskip\noindent
where $V(\phi_1,\phi_2,\phi_3)$ is the initial potential in the Jordan frame
and  $z_j=\phi_j/\rho$ are functions of $\theta$, $\zeta$ only.
If  $\xi_1=\xi_2=\xi_3$, then
the Planck scale is also determined by the same $\rho$ field.
The extension to more scalar fields  is straightforward. This study can also be extended to include 
additional (Weyl gauge invariant) terms  quadratic in the scalar curvature \cite{new1,new2}.

\subsection{Other implications: UV scale-invariant regularization}\label{3.4}

The above results  have  implications for
models with (global)  scale  invariance at the quantum level. Such
models are important since they can have a  quantum stable hierarchy between two scalar
fields vev's (higgs and dilaton), which is  relevant for the SM hierarchy problem, as we detail below.

Consider first a classical scale invariant model. The SM with a vanishing higgs mass parameter is an
example. This symmetry can  be preserved at the quantum level, by ensuring  
that the UV regularization  respects it. 
This is done by replacing the subtraction scale $\mu$ by the dilaton 
field  $\rho$ \cite{Englert}.  After spontaneous
breaking of this symmetry, $\mu\sim \,\langle\rho\rangle$.  In this way one 
obtains scale invariant results at the quantum  level \cite{S1,Armillis,D1,D2,Monin}. After the quantum 
calculation one can expand the result (e.g. the scalar potential) about the vev of the
dilaton to recover standard results (e.g. Coleman-Weinberg potential) plus additional
higher dimensional operators suppressed by the dilaton vev \cite{D1}. Such models have 
only spontaneous breaking  of the scale symmetry,  thus there is no dilatation   anomaly  
\cite{Englert,Armillis,Tamarit,D2}.

The relation to the hierarchy problem is that the dilaton vev is fixing $M_p$ 
and so it must be much higher than the Higgs vev. Such hierarchy can be the result of 
 one initial {\it classical} tuning of the (dimensionless) couplings. 
This tuning remains stable at the quantum level, due to quantum scale 
invariance and a shift symmetry of the dilaton (Goldstone mode)~\cite{K1}.
However, the dilaton remains in the spectrum  as a flat direction, 
even  at  quantum level. {One can then ask  what happens
to this flat direction for a  more general, local Weyl symmetry.}

The result of this paper answers this question.  As we saw, the  dilaton is ``eaten'' 
by the Weyl field $\w_\mu$ which becomes massive, decouples from the spectrum and
{leaves a potential   function of the angular variables fields 
only (which can be the Higgs field, inflaton, etc).}

{
For example, in a two-field case and assuming a potential $V$ of eq.(\ref{Vh2}), 
after decoupling the potential $\hat V$
depends only on $h\equiv v\cot\theta$ which can be  the neutral Higgs field;
taking for simplicity $\xi_2=0$ and large $\xi_1$, after dilaton decoupling,
the scalar potential (for a canonical kinetic term for $h$) becomes in the Einstein frame
\medskip
\bea
\hat V=\frac{\lambda_2}{4!} 
\Big( \,h^2- \frac{\sqrt\lambda_1}{\sqrt\lambda_2}\frac{v^2}{(\xi_1/6)}\Big)^2,
\eea}

\medskip\noindent
{which is indeed that  of the SM, with $m_h^2=\sqrt{\lambda_1\lambda_2}\,v^2/\xi_1
\!\ll\! v^2\sim \langle\rho\rangle^2 $ 
for small (ultra-weak) couplings and large $\xi_1$. 
What happens with this hierarchy at the quantum level? }

{Before the Stueckelberg mechanism, the dilaton can enforce a scale (or Weyl) invariant 
UV regularisation \cite{Englert}
of the quantum corrections to potential (\ref{Vh2}), as described above.
In this way one can construct a quantum scale (Weyl) invariant theory, 
dilatation (conformal) anomaly-free, respectively.
 In this case the mentioned classical hierarchy
 between $m_h^2$ and $v^2\sim \langle\rho\rangle^2$  remains stable at the  quantum level 
(for more details  see discussion in \cite{D1,D2}).}

\section{Inflation from Weyl gauge symmetry}\label{4}

In this section we study  inflation in  models with spontaneously  broken Weyl gauge 
symmetry.  We consider the case of two scalar fields of Section~\ref{section3.2}
 and regard the potential for the angular-variable field $\theta$, obtained after 
the Stueckelberg mechanism,  as being responsible for inflation. The potential becomes
constant at large $\tan\theta$.  
In this limit, from eqs.(\ref{Lfinal}), (\ref{thetaf}), $\theta$ and $\w_\mu^\prime$ decouple 
 and the action for the inflaton ($\theta$)  can be written as 
%\medskip
\bea
L_{\rm infl}
=  \sge \Big[-\frac{1}{2}\,v^2\, \RE  +  \frac{3b\,v^2}{\xi_2}\, \frac{\tau^2
+\xi_2/\xi_1}{(1+\tau^2) (\tau^2+b)^2}\, (\partial_\mu\tau)^2-\VE(\tau)\Big],
\qquad
\tau\equiv \tan\theta 
\label{inflaton}
\eea

\medskip\noindent
For $V$ of (\ref{Vh1}), the Einstein-frame potential  expressed in terms of $\tau$ is:
\medskip
\bea
\VE &=&\frac{36\,v^4\,}{(\xi_1\phi^2_1+\xi_2\phi^2_2)^2}\, \Big(\frac{\lambda_1}{4!}
\,\phi_1^4+\frac{\lambda_{12}}{4}\,\phi^2_1\,\phi^2_2
+\frac{\lambda_2}{4!}\,\phi_2^4 \Big)  
\nonumber \\
&=&\frac{36\,v^4\, (1+\xi_1)^2}{\xi^2_1(\tau^2+b)^2} \, 
\Big(c_1 \tau^4+c_{12}\tau^2+c_2 \Big), \label{inflatonpot}
\eea
with
\bea
c_1= \frac{\lambda_1}{4!(1+\xi_1)^2}, \qquad
c_{12} = \frac{\lambda_{12}}{4(1+\xi_1)(1+\xi_2)}, \qquad
c_2 = \frac{\lambda_2}{4!(1+\xi_2)^2}.
\eea

\medskip\noindent
If $c_{12}=-2\sqrt{c_1 c_2}$ ($c_{1,2}>0$), then $\hat V=0$ at the minimum.
The potential is similar to that in  Higgs portal inflation \cite{lebedev,higgsdilaton}, 
but in our case  the angular field $\theta$ is the dynamical field responsible for a slow-roll inflation 
 (instead of being frozen). For the case of  global Weyl invariant models inflation was
already studied in \cite{FRH2,GG1}.

In the following we consider the case   $\xi_1=\xi_2$, 
since then $\theta$ and $\w_\mu^\prime$ are 
decoupled from each other for all values of $\theta$ (c.f.(\ref{Lfinal})). Then  $b=1$,  and  the kinetic
term acquires a canonical form  $(1/2)\,(\partial_\mu\chi)^2$,
 with  the  actual   inflaton  field $\chi$  defined by
\medskip
\bea
\chi
%= v \sqrt{\frac{6}{\xi_1}}\, \arctan(\tau)
=v\,\theta \,\sqrt{\frac{6}{\xi_1}}.  \label{canon}
\eea

\medskip\noindent
Then the inflaton potential in eq.~(\ref{inflatonpot}) becomes
\medskip
\bea
{\hat V}=
V_0 \cos^4\theta\Big(\tan^4\theta+\frac{c_{12}}{c_1}\,\tan^2\theta+\frac{c_2}{c_1} \Big),
\quad\text{with}\quad
V_0=\frac{36\, v^4}{\xi^2_1}\,c_1\,(1+\xi_1)^2 \,
\eea

\begin{figure}[t!]
   \begin{center}
     \includegraphics[height=0.40\textwidth]{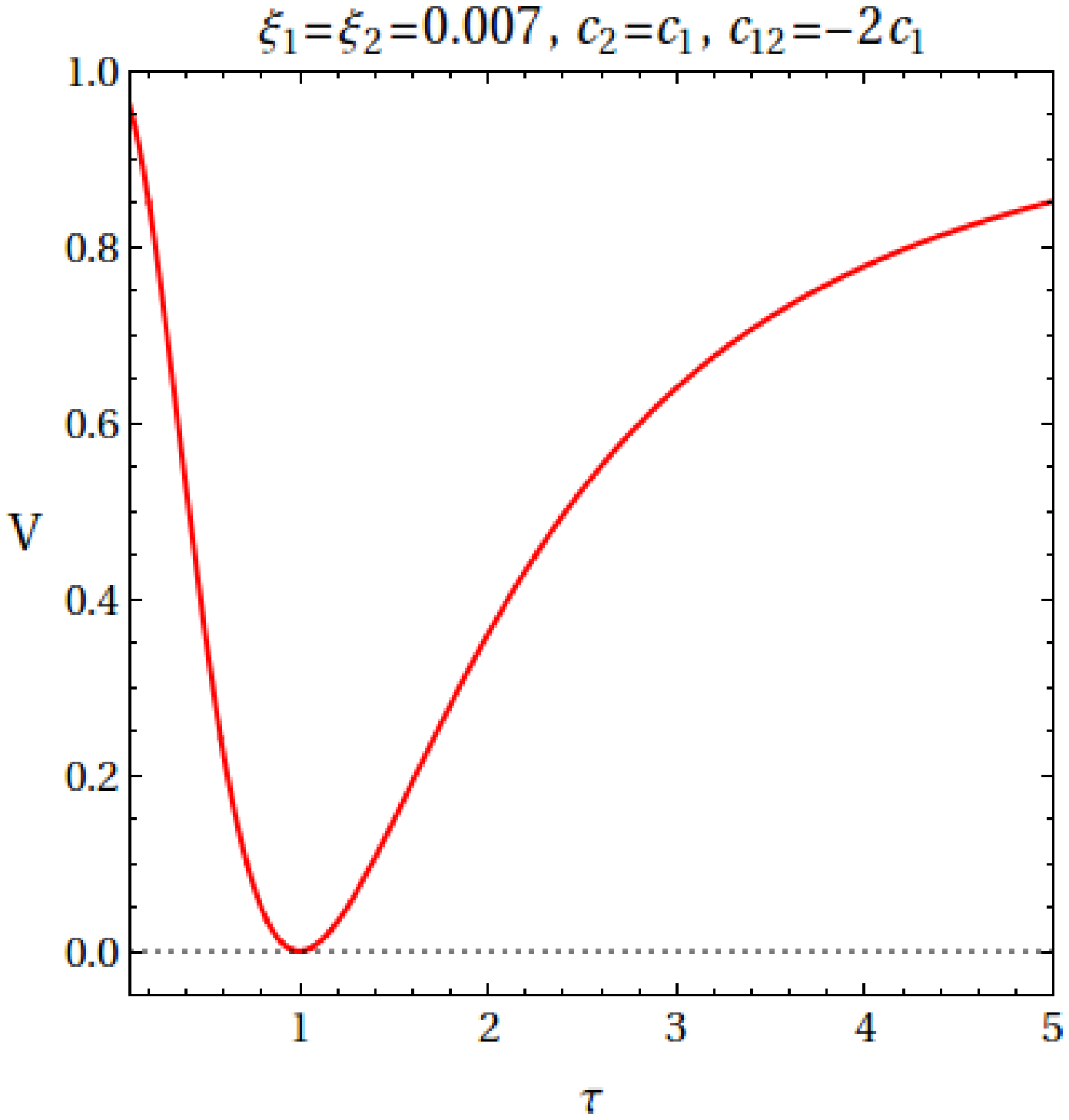} \quad       
     \includegraphics[height=0.40\textwidth]{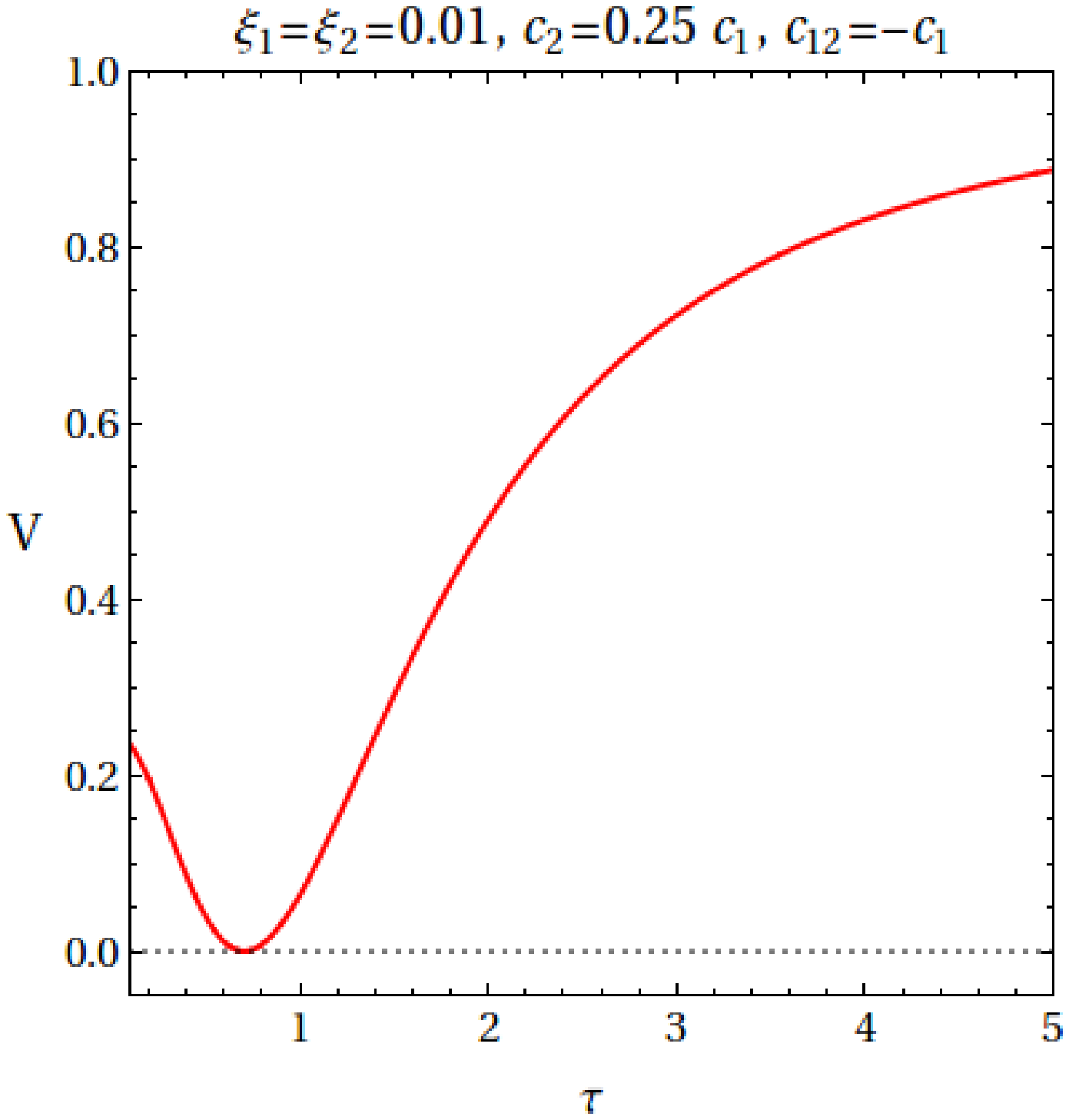} 
       \end{center}
  \caption{\small Inflaton potential (divided by constant $V_0$) as a function of $\tau=\tan\theta$,
  for $\xi_1=\xi_2$ and  ratios  $c_{12}/c_1$, $c_2/c_1$ of values shown above.
  $V=0$ at the minimum;
note $\theta=(\chi/v)\sqrt{\xi_1/6}$ with $\chi$ the actual inflaton. The 
spectral index and tensor-to-scalar ratio for these  cases are shown in Figure~\ref{plot}.}
  \label{pot}
\end{figure}

\medskip\noindent
 with $\theta\equiv\sqrt{\xi_1/6}\,\chi/v $.  The potential  is illustrated 
in Figure~\ref{pot} for some choices of the quartic couplings, with  $\hat V=0$ at the minimum.
Inflation takes place at $\tan\theta\gg 1$  (or $\theta\sim\frac{\pi}{2}$)
for which $V=V_0\, [1+ (c_{12}/c_1-2)\cot^2\theta +(3 -2 \,c_{12}/c_1 +c_2/c_1)\cot^4\theta+\cdots]$.
 For $\xi_1\neq \xi_2$ with $\tan\theta\gg 1$ 
 we note that there is an approximate relation between $\theta$ 
 and the canonical inflaton field $\chi$  similar to that in eq.~(\ref{canon}).

The slow-roll parameters (for $\xi_1=\xi_2$) are given by
\medskip
\bea
\epsilon &=& \frac{1}{3}\,\xi_1 \,\frac{\Big(c_2-c_1+(c_1-c_{12}+c_2)\cos(2\theta)\Big)^2\sec^4\theta\tan^2\theta}{(c_2+c_{12}\tan^2\theta+c_1\tan^4\theta)^2}, \label{eps} \\
\eta&=& -\frac{1}{3}\, \xi_1\, \frac{\Big((c_2-c_1)\cos(2\theta)+(c_1-c_{12}+c_2)\cos(4\theta)\Big)\sec^4\theta}{c_2+c_{12}\tan^2\theta+c_1\tan^4\theta}.  \label{eta}
\eea

\medskip\noindent
We find that $\epsilon$ and $\eta$  can be small simultaneously for $\xi_1\ll 1$. 
 \begin{figure}[t!]
  \begin{center}
    \includegraphics[height=0.40\textwidth]{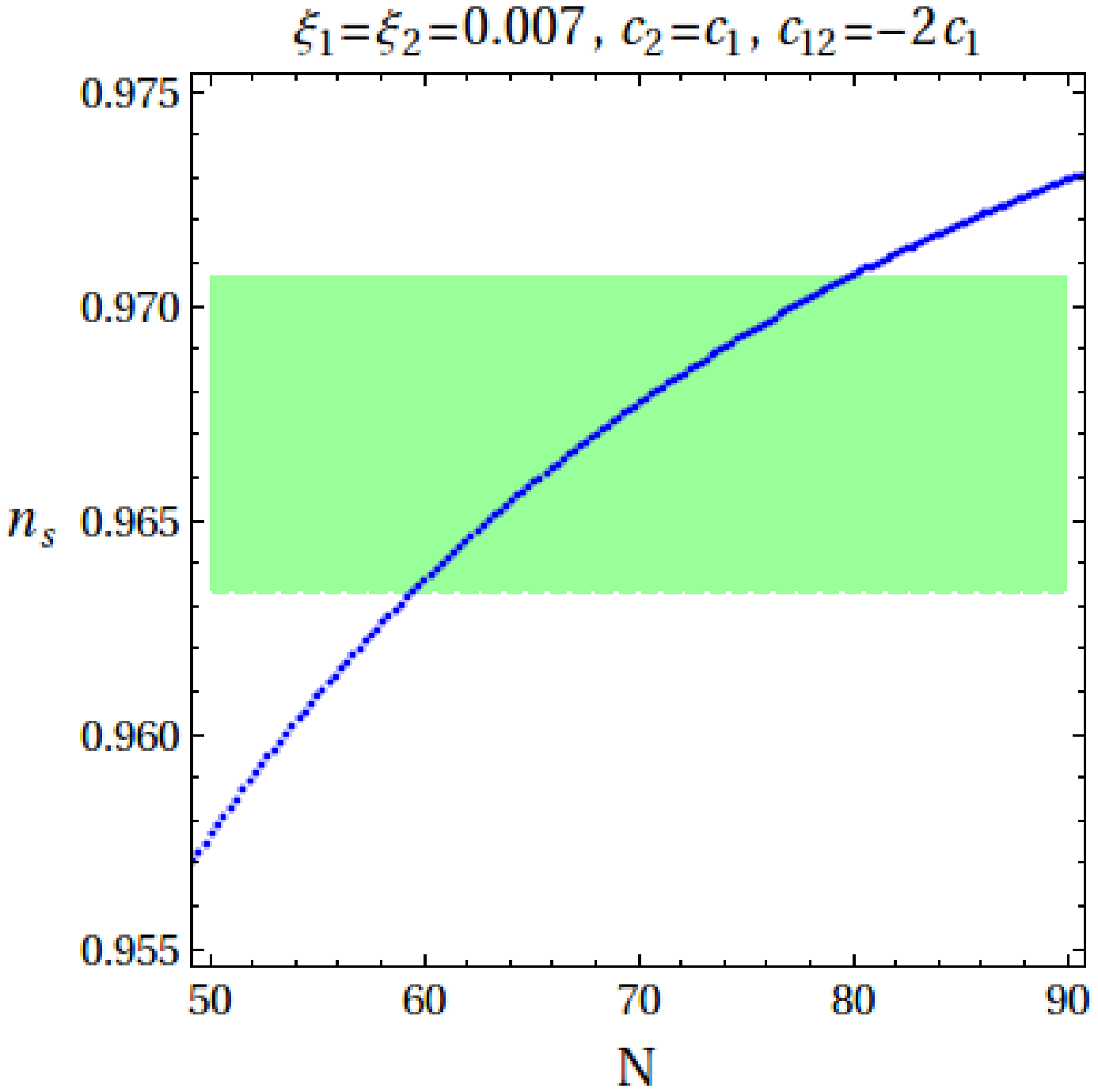} \quad       
     \includegraphics[height=0.40\textwidth]{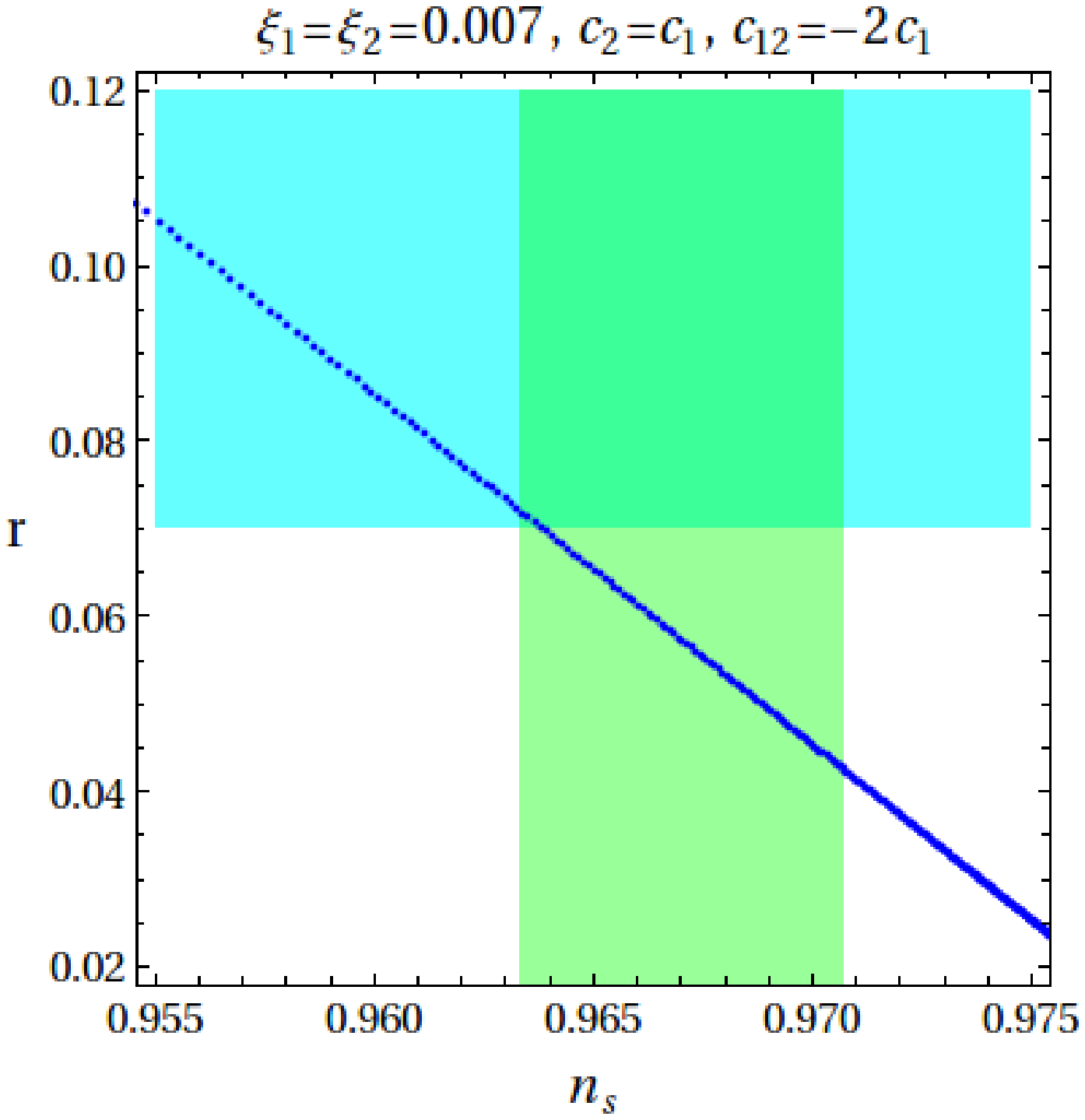} \\
    \includegraphics[height=0.40\textwidth]{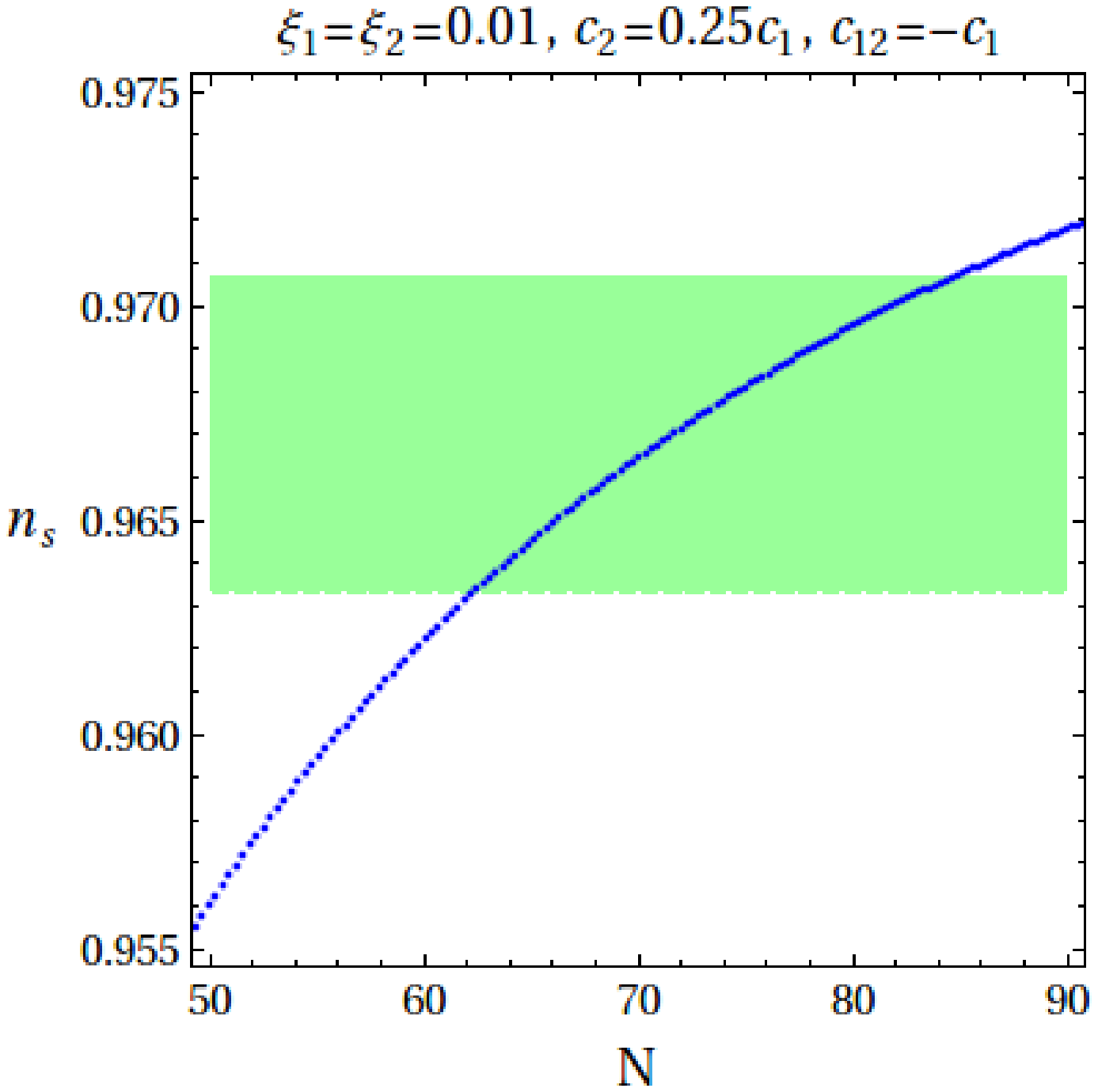} \quad       
     \includegraphics[height=0.40\textwidth]{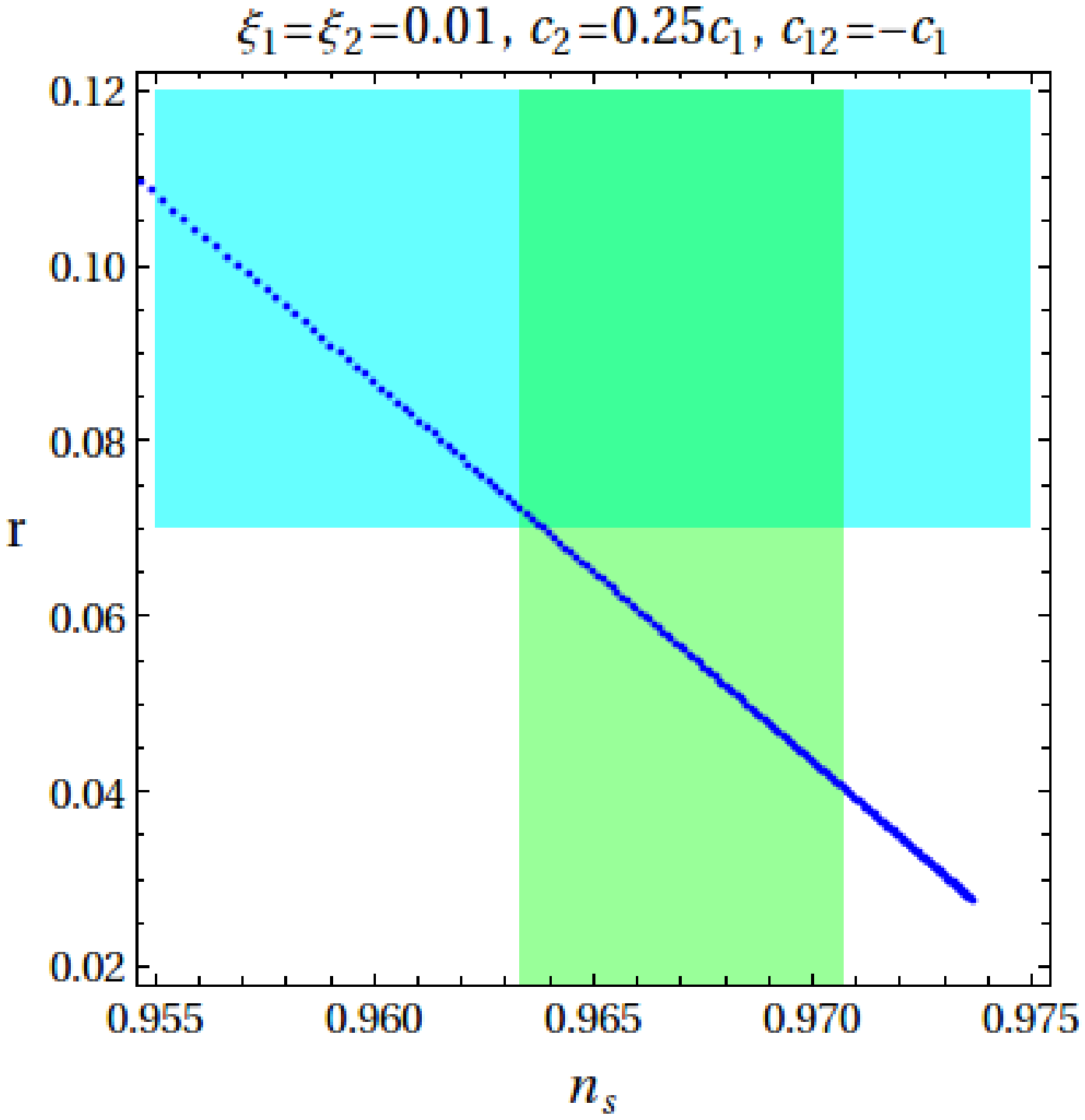} 
  \end{center}
  \caption{\small{
Left: The spectral index $n_s$ as a function of the number of e-foldings $N$ for two different
cases. 
Right:  The spectral index $n_s$ versus tensor-to-scalar ratio $r$ for the corresponding cases. 
The green region in 
all the plots corresponds to Planck2018 values for $n_s$ within $1\sigma$ (see text).
The cyan region with $r>0.07$ 
is excluded by Planck at 95\% CL. We chose $\xi_1=\xi_2$ for all the plots and 
$c_2/c_1=1$ and $c_{12}/c_1=(-2)$ (upper panels) and $c_2/c_1=0.25$ and $c_{12}/c_1=(-1)$
(lower panels). Here $c_{12}$ is fixed by the condition ${\hat V}=0$ at the minimum. 
 }}
  \label{plot}
\end{figure}

If we choose $c_2=c_1$, $c_{12}=-2 c_1$ then the expressions of the 
slow-roll  parameters simplify further:
\bea
\epsilon &=& \frac{4}{3} \, \xi_1 \tan^2(2\theta),  \label{eps1} \\
\eta&=& \frac{4}{3} \, \xi_1\Big(-1+\tan^2(2\theta) \Big),
\eea

\medskip\noindent
which leads to the scalar spectral index and the tensor-to-scalar ratio as
\medskip
\bea
n_s&=&1 +2\,\eta_*
-6\epsilon_*= 1-\frac{8}{3}\xi_1 \Big(1+2\tan^2(2\theta_*) \Big),
\eea
and
\bea
r&=& 16\,\epsilon_* =\frac{64}{3} \,\xi_1 \tan^2(2\theta_*).
\eea

\medskip\noindent
Further, the  number of e-foldings  during inflation is also given by
%\medskip
\bea
N&=&  \, v^{-1}\int^{\chi_*}_{\chi_{\rm end}} \frac{{\rm sign}({\hat V}') d\chi}{\sqrt{2\epsilon(\chi)}} \nonumber \\
&=& \frac{3}{4\xi_1}\, \log\bigg|\frac{\sin(2\theta_{\rm end})}{\sin(2\theta_*)}\bigg|
\eea

\medskip\noindent
where $\theta_*$ is evaluated at the horizon exit and $\theta_{\rm end}$ is the inflaton 
value at the end of inflation. Inflation ends at $\epsilon=1$, i.e. $|\tan(2\theta_{\rm end})|=(\frac{3}{4\xi_1})^{1/2}$ from eq.~(\ref{eps1}).

The normalization of the CMB anisotropies,
$V_0/(24\pi^2 v^4\,\epsilon_*)=2.1\times 10^{-9}$  \cite{planck2018}, constrains $c_1$ 
(or equivalently the quartic coupling $\lambda_1$) and the non-minimal
 couplings $\xi_{1,2}$ to satisfy
%\medskip
\bea
\frac{c_1(1+\xi_1)^2}{\xi_1^3}= 1.8\times 10^{-8} \tan^2(2\theta_*).
\eea
This constraint is respected  by choosing small values  of $c_1$ (or $\lambda_1$),
for given $\xi_1$.

We have that $n_s=0.9670\pm 0.0039$ ($68\%$ CL) and $r < 0.07$ ($95\%$ C.L.)
 from Planck 2018 (TT, TE, EE + low E + lensing + BK14 + BAO) \cite{planck2018}.
In Figure~\ref{plot}, we illustrated the relation between the spectral index versus
 the number of e-foldings in the left plots and showed the spectral index versus the 
tensor-to-scalar ratio in the right plots. Here we have fixed $\xi_1=\xi_2=0.007$,
 $c_2=c_1$ and $c_{12}=-2c_1$ in the upper panel, and $\xi_1=\xi_2=0.01$, $c_2=0.25c_1$,
 $c_{12}=-c_1$ in the lower panel. As a result, we find  that our model of inflation 
is consistent 
with the observed spectral index and the bound on $r$, for small non-minimal couplings.

\section{Conclusions}\label{5}

In this work we discussed the Weyl conformal symmetry and its spontaneous breaking 
and some implications for model  building beyond the SM and inflation.

In models with conformal symmetry (of the Brans-Dicke-Jordan type) 
with scalar fields with non-minimal couplings to the Ricci scalar,  one can generate 
spontaneously the Planck scale from the vev of a scalar field (or a combination of them). 
However, a positive (negative) Newton constant is accompanied by a negative (positive) kinetic 
term  for this field, respectively. 
This situation  is naturally avoided in models 
with an  additional Weyl gauge symmetry and a gauge field $\w_\mu$ which is of geometric origin,
 with a gauge transformation  dictated by  the conformal  transformation of the metric.

We showed that the Weyl field $\w_\mu$ couples only to the scalar sector but not to the fermionic
sector of a SM-like Lagrangian in curved space-time, which  is interesting for model building.
Further, the field $\w_\mu$ undergoes a Stueckelberg mechanism and 
becomes massive after ``eating'' the radial mode $\rho\sim \sqrt K$  (in field space) 
and would-be-Goldstone mode (dilaton). 
The Weyl gauge symmetry is then spontaneously broken (and there are no negative kinetic
terms  in the theory). Further, the  vev $\langle\rho\rangle$ 
determines the mass of  $\w_\mu$ and the Planck scale $M_p$ (up to possible
additional angular-variables field dependence). The mass of $\w_\mu$ can be larger or
smaller than $M_p$ depending on  the scalar fields charge and non-minimal couplings.
 After decoupling  of $\w_\mu$ the potential  depends on the angular variables fields only
which can play the role of the neutral Higgs field, inflaton, etc.
For two scalar fields of equal non-minimal couplings,
the field $\w_\mu$   decouples  from the action even if it is light.

For the case with two scalar fields, the scalar potential generally depends only on the angular variable
 field $\theta$, and it is nearly constant at large $\tan\theta$, when $\w_\mu$ also decouples. 
 Therefore, the potential can be relevant  for a single-field inflation. 
Investigating the details of the inflaton potential, we found that successful inflation 
is possible, with values of $n_s$ and $r$ consistent with Planck2018 constraints,
  for perturbative values of the couplings.

While this  study was  formulated in (pseudo)Riemannian  geometry extended with a 
real Weyl field (undergoing a gauge transformation dictated by conformal transformation 
of the metric), 
the natural framework is that of Weyl conformal geometry where this symmetry is manifest.
In the Riemannian case imposing this symmetry avoids the ghost kinetic term of conformal theory
and leads to a Lagrangian with a current $\partial_\mu K$ that interacts with the Weyl field.
This  Lagrangian was shown to be  identical, up to a total derivative term, to that obtained in 
Weyl geometry (WG) where the Weyl symmetric Lagrangian is naturally
built-in,  with  curvature scalar, tensors  and affine connection of Weyl geometry. 
This equivalence is showed for a SM-like Lagrangian endowed with Weyl gauge symmetry,
using the relation between $R$ computed in Riemann geometry 
with Levi-Civita connection and its  counterpart in Weyl geometry.
This Lagrangian can be used for examining the phenomenological constraints on the
SM extended with Weyl gauge symmetry.

\vspace{1cm}
\noindent
{\bf Acknowledgements:\,\,\,\,} 
The authors are grateful for hospitality and  support to CERN Theory Division
where this work was done. 
The work of H.M.L. is supported in part by Basic Science Research Program through the National 
Research Foundation of Korea (NRF) funded by the Ministry of Education, 
Science and Technology (NRF-2016R1A2B4008759 and NRF-2018R1A4A1025334). 
D.M.G. is supported by National Programme PN 18090101.

\end{document}